# In-Plane Magnetic Anisotropy In RF Sputtered Fe-N Thin Films

H. B. Nie[1], S. Y. Xu[1], C. K. Ong[1], Q. Zhan[2], D. X. Li[2] and J. P. Wang[3]

[1] *Centre for Superconducting and Magnetic Materials, Institute of Engineering Science and Department of Physics, National University of Singapore, 2 Science Drive 3, Singapore 117542.*

[2] *Shenyang National Laboratory for Materials Science, Institute of Metal Research, Chinese Academy of Sciences, 72 Wenhua Road, Shenyang 110016, China.*

[3] *Media & Materials, Data Storage Institute, DSI Building, 5 Engineering Drive 1, Singapore 117608*

**Abstract**

We have fabricated Fe(N) thin films with varied $N_2$ partial pressure and studied the microstructure, morphology, magnetic properties and resistivity by using X-ray diffraction, atomic force microscopy, transmission electron microscopy, vibrating-sample magnetometer and angle-resolved M-H hysteresis Loop tracer and standard four-point probe method. In the presence of low $N_2$ partial pressure, Fe(N) films showed a basic *bcc* $\alpha$-Fe structure with a preferred (110) texture. A variation of in-plane magnetic anisotropy of the Fe(N) films was observed with the changing of N component. The evolution of in-plane anisotropy in the films was attributed to the directional order mechanism. Nitrogen atoms play an important role in refining the $\alpha$-Fe grains and inducing uniaxial anisotropy.

## 1   Introduction

To write data on high coercivity media, high saturation magnetization and low coercivity are required for inductive head materials. An addition of a small amount of nitrogen to a pure film leads to a significant improvement of soft magnetic properties of the iron film [1-4]. Grain size and the lattice dilation due to interstitial nitrogen atoms are important facts in controlling the properties of nanocrystalline Fe-N films [5, 6]. High magnetization and soft magnetic properties have made FeN-based films very attractive for inductive heads [1, 7, 8]. Additionally, well-defined uniaxial anisotropy is also important for a head material [3, 9]. Magnetic anisotropy must be thoroughly understood before it can be implemented in magnetic recording applications [10, 11]. The stress in thin films can be largely influenced by deposition conditions, such as gas pressure [12, 13]. Nagai *et al.* [3] studied anisotropy in iron film formed by ion beam sputtering, and argued that the uniaxial anisotropy in those films was mainly caused by directional ordering rather than crystalline anisotropy. Takahashi *et al.* [5] analyzed magnetic anisotropy dispersion in Fe based alloy films. They demonstrated that the reduction of grain size and induced lattice deformation were very effective for increasing the value of initial permeability. Morisako *et al.* [14] reported magnetic anisotropy of iron nitride thin films prepared by facing-target sputtering, they found that the magnetic anisotropy was induced into the films at nitrogen partial pressure above 0.07 mTorr and the direction of the easy axis was parallel to the magnetic stray field. Viala *et al.* [6] showed local magnetic anisotropy in nitrogented iron-based soft magnetic films. In the present work, we synthesized Fe-N thin films using radio frequency (RF) reactive sputtering under various $N_2$ partial pressures ($P_N$), and studied the structural and magnetic properties of the films. The in-plane anisotropy of the films was studied by means of angle-resolved M-H loop tracer. A variation of in-plane magnetic anisotropy of the films was clearly observed through the change of hard-axis squareness $S_d$ of the hysteresis loops. The origin of the variation of in-plane magnetic anisotropy is discussed.

## 2. Experimental procedures

The Fe-N thin films were fabricted by using reactive RF magnetron sputtering deposition in a Denton Vacuum Discovery-18 Deposition System with a 3-inch-diameter iron target (purity 99.9%). The substrate stage was placed horizontally with a 30° angle to the target surface. The distance between the center of the target and the substrate stage is 11 cm. Si(100) substrates were chemically cleaned with a standard process [15]. The system was first pumped to $8.0 \times 10^{-7}$ torr, and then a mixture of pure Ar/$N_2$ gases (purity 99.9995%) was introduced in the deposition chamber through mass flow controllers. The total processing ambient gas pressure was kept at 2.0 mtorr for all the films, while the relative $N_2$ partial pressure $P_N$ was varied from 1.0% up to 12.0% for a series of Fe-N films. The target and substrate surfaces were pre-sputtered before each deposition. The RF power was kept at 100 W, corresponding to a power density of 2.2 W/cm$^2$ on the target surface. During the sputtering process, an aligning magnetic field $\bar{H}_{al}$ around 150 Oe induced by permanent magnets was applied parallel to the substrate surface, and a –100 V dc bias voltage was applied to the substrate stage which was rotated at 9 RPM. The permanent magnets were fixed on the substrate stage so that the direction of the aligning magnetic field $\bar{H}_{al}$ with the substrates was fixed. Each film was deposited for 10 minutes, corresponding to a film thickness of around 30 nm.

Without performing any annealing processes on the samples, we characterized the as-deposited samples using X-ray diffraction (XRD), atomic force microscopy (AFM), transmission electron microscopy (TEM), angle-resolved M-H loop tracer, vibrating-sample magnetometer (VSM) and four-probe method to study their structural and magnetic properties. For XRD measurements, the author used both continuous $\theta - 2\theta$ scan and step $\theta - 2\theta$ scan mode. During our measurements, a generator tension of 30 kV and a current of 20 mA were used. For continuous $\theta - 2\theta$ scan, the author selected scanning time of 1 second per step and the step size of 0.02°. For the step $\theta - 2\theta$ scan, the author selected scanning time of 2.5 second per step and the step size of 0.005°. We used AFM tapping$^{TM}$ mode to examine the surface structures of the thin films, and silicon cantilevers were used. The image obtained is full 16-bit on all axes. The resonance frequency of the cantilever was about 260 KHz, and the scan rate is around 0.4 Hz. The specimens suitable for TEM observations were prepared by standard techniques. Plan-view specimens were made parallel to the Si (100) plane and thinned from the substrate side. The specimens were examined in a JEM-2010 high-resolution electron microscope with a point-to-point resolution of ~ 0.19 nm. To reveal the magnetic anisotropy of the films, we studied the evolution of *M-H* hysteresis loops as a function of angle ϕ using the M-H loop tracer under the maximum field of 100 Oe. Where ϕ is the angle between the directions of the applied magnetic field $\vec{H}_m$ for M-H loop measurement and the easy axis in the film. Each sample was measured under varied angle ϕ from 0$^o$ to 360$^o$, in 15$^o$ increments. Magnetization under an applied field of 1000 Oe was measured with VSM. The thickness of the films was measured by using an Alpha-step 500 surface profiler.

## 3. Results and discussions

We determined the crystalline structures of the Fe-N thin films by using normal *θ-2θ* XRD scans. We found that the pure Fe film shows the (110) peak of the *bcc* (body-centered cubic) α-Fe structure. When $N_2$ was introduced in the deposition process, at a low $N_2$ partial pressure, $P_N$, *bcc* crystalline structure was found dominating in the Fe(N) films, where Fe(N) representing that part of nitrogen atoms have taken the interstitial position of the *bcc* Fe lattice. At $P_N$ of smaller than 8.0% the films showed mainly α-Fe(N)(110) peaks, indicating that the main phase in these films was nanocrystalline α-Fe(N). When $P_N$ was lager than 8.0%, almost no detectable α-Fe(N)(110) peak appeared, indicating that the as-deposited films became almost amorphous. With the changing of $P_N$, the position shift of the α-Fe(N)(110) peaks in the curves of XRD figures was noted, so we carefully performed step scans (see figure 1(a)) for each sample and calculated their corresponding lattice constants. We found that with increasing $P_N$, α-Fe(N) (iron) (110) peak shifts toward lower angle direction, which indicated an enlarged lattice perpendicular to the plane of the film (figure 1(b)) or an increasing compressive stress component. Since Fe-N is a defect compound system where a wide atomic percentage range of N is allowed in its stable phases, lattice distortions can be expected in each phase. The N atom is so small that it can fit into the holes between the iron atoms, either at the centers of the edges or at the centers of the faces of the unit cell [10]. There are interstitial nitrogen atoms distributed in the α-Fe(N) lattices in Fe-N thin films. Since the lattice distance $d_{(110)}$ increased with the increasing of $P_N$, we may assume that the N atoms preferentially occupy the interstitial sites in iron in a plane normal to the local magnetization, namely, in a plane normal to the aligning magnetic field $\vec{H}_{al}$. This consists with the C atoms preferentially occupy the interstitial sites in iron in the presence of an applied field [10]. From figure 1(a), we got that α-Fe(N) (110) peak intensity increased with increasing $P_N$ from 0.0% to 4.0%, then decreased with further increasing $P_N$. It indicated that the amount of nanocrystalline α-Fe(N) increased first, then decreased. With increasing $P_N$, the α−Fe(N) (iron) (110) peak became broader, which indicating that the nanocrystalline α-Fe(N) grain size decreased with increasing $P_N$ (see fig. 4 (a)).

Figure 2(a) shows dependence of magnetic easy (φ=0°) axis coercivity $H_c^e$ and magnetic hard (φ=90°) axis coercivity $H_c^d$ of the samples on $P_N$. The coercivity $H_c$ has minimum values of 2-3 Oe when $P_N$ ranges in 3.0% to 7.0%. On the contrary, with the increase of $P_N$, the saturation magnetization $M_s$ of the films decreased (fig. 2(a)), due to a decreasing of amount of high moment ferromagnetic Fe and an increasing amount of non-magnetic nitrogen content in the samples. However, with $P_N$ from 0.0% to 7.0%, $M_s$ decreased slowly, but with $P_N$ from 8.0% to 12.0%, $M_s$ decreased quickly. Both of them could be explained that, when the nitrogen partial pressure $P_N$ is low, nanocrystalline α-Fe(N) is still dominating in the films; while further increase of $P_N$, the amorphous structure dominate the films. This can be seen from the XRD patterns (fig 1(a)), when $P_N$ is larger than 8.0%, almost no α-Fe(N) (110) peak can be detected. Low saturation magnetization and high coercivity consistent with the amorphous structure of the films.

The in-plane *M-H* hysteresis loops of the samples showed that all the films had their easy axes lying along the direction of aligning magnetic field $\vec{H}_{al}$. With the increase of $P_N$, from 0.0% to 12.0%, the easy-axis squareness S(0°) or $S_e$ ($S = M_r/M_s$) of the in-plane *M-H* hysteresis loops is almost a constant but the hard-axis squareness S(90°) or $S_h$ shows a similar trend of $H_c$ and minimizes at 0.20 when $P_N$ = 7.0% (fig. 2(b)). Note that pure Fe film is almost isotropic in the film plane. With nitrogen atoms introducing into the pure Fe film, in-plane uniaxial anisotropy was found. For Fe(N) films, N atoms enter α-Fe as interstitial atoms and expend the *bcc* crystalline lattice, so the cubic symmetry of α-Fe is broken and uniaxial anisotropy is induced [16, 17] With the increase of $P_N$ from 0.0% to 7.0%, the in-plane uniaxial anisotropy increases, showing a S(0°) of 0.96 and S(90°) of 0.20. Further increasing $P_N$ from 7.0% to 11.0%, the uniaxial anisotropy decreased greatly. When $P_N$ was 12.0%, there was almost no uniaxial anisotropy in the film. Obviously, here the nitrogen content in the film plays a key role. The phenomenon consists with the XRD patterns we obtained. With increasing $P_N$ from 7.0% to 12.0%, α-Fe(N) (iron) (110) peak becomes broader and weaker, and disappears finally, indicating a large amount of disorders in the α-Fe(N) lattices that suppress the uniaxial anisotropy.

Figure 3 shows polar diagrams of coercivities as a function of angle φ, and Easy (φ=0°) and hard (φ=90°) axis hysteresis loops of Fe-N thin films when $P_N$=0.0% (a), 3.0% (b) and 7.0% (c), respectively. When $P_N$=0.0%, there was almost no anisotropy in the film plane. When $P_N$=3.0%, in-plane uniaxial anisotropy was observed, and the strongest in-plane uniaxial anisotropy was shown at $P_N$=7.0%. As the interstitial N atoms play a key role in controlling the magnetic anisotropy in Fe-N films, the main reason that the variation of in-plane magnetic anisotropy may be attributed to directional order mechanism [10, 18]. As the N atoms preferentially occupy the interstitial sites in iron in a plane normal to the local magnetization, we may assume that some pairs of iron and/or iron-nitrogen particles are statically ordered under the aligning magnetic field $\vec{H}_{al}$ during film deposition [3]. For pure Fe film, iron atoms are homogeneously disordered, there is no directional order, so the pure Fe film is isotropic, and $S_e \approx S_h$. With a little nitrogen atoms introducing into the pure Fe film, some pairs of iron and/or iron-nitrogen particles are statically ordered under the $\vec{H}_{al}$ during film deposition, in-plane uniaxial anisotropy was induced, and the easy axis was along the direction of the $\vec{H}_{al}$. The strongest in-plane uniaxial anisotropy was obtained at $P_N$=7.0%. In this case the directional order was strongest. When $P_N$=12.0%, the amorphous structure dominate the films, a large amount of disorders in the α-Fe(N) lattices suppress the uniaxial anisotropy, so the film became to be isotropic, and $S_e \approx S_h$.

Figure 4(a) shows average grain sizes calculated from the XRD step-scan patterns according to Scherrer equation as a function of $P_N$. The grain size decreased with increasing of $P_N$. To reveal the grain size of the film, we also investigated the plane view of the sample. Figure 5 is a TEM dark-field plane-view of the film when $P_N$=5.0%. The grains have irregular shapes with grain size of about 12 nm. Fig. 6 shows the AFM tapping mode micrographs of the Fe(N) thin films, corresponding to $P_N$ of 0.0%, 3.0%, and 12.0%, respectively. It clearly shows that grain size decreased with nitrogen added, and the film has amorphous structure when $P_N$ = 12.0%. The resistivity ρ of the as-deposited films (figure 4(b)) was determined by using standard four-point method at room temperature. The resistivity ρ increased with the increasing of $P_N$ from 0.0% to 12.0%. The resistivity ρ is high (between 30 and 220 μΩ•cm). With the increasing of $P_N$, there is more non-metallic nitrogen content in the film. This will cause the number of free electrons per unit volume decreasing. And it will create more defects (dislocations, porus, impurities, strain) and disorder of atom arrangement in crystallites also. The second factor is that with $P_N$ increasing, grain size decreased, the grain boundary scattering is remarkably enhanced. Therefore the resistivity ρ increase with the increasing of $P_N$. High electrical resistivity of the films when $P_N$ is larger than 8.0% is partly due to the amorphous phase. High values of resistivity ρ can reduce eddy current loss. In this respect, using Fe(N) thin films in magnetic recording head is a good choice.

## 4. Conclusions

We have fabricated Fe(N) thin films with varied $N_2$ partial pressure. Fe(N) films showed a basic *bcc* α-Fe structure with a preferred (110) texture. A variation of in-plane magnetic anisotropy of the films was clearly observed through the change of hard-axis squareness $S_h$ of the hysteresis loops with the changing of N component. This variation is accompanying with a similar variation of textural structure observed in the samples. The evolution of in-plane uniaxial anisotropy in Fe(N) thin films was attributed to the directional order mechanism. Nitrogen atoms play an important role in refining the α-Fe grains and inducing uniaxial in the Fe(N) films.

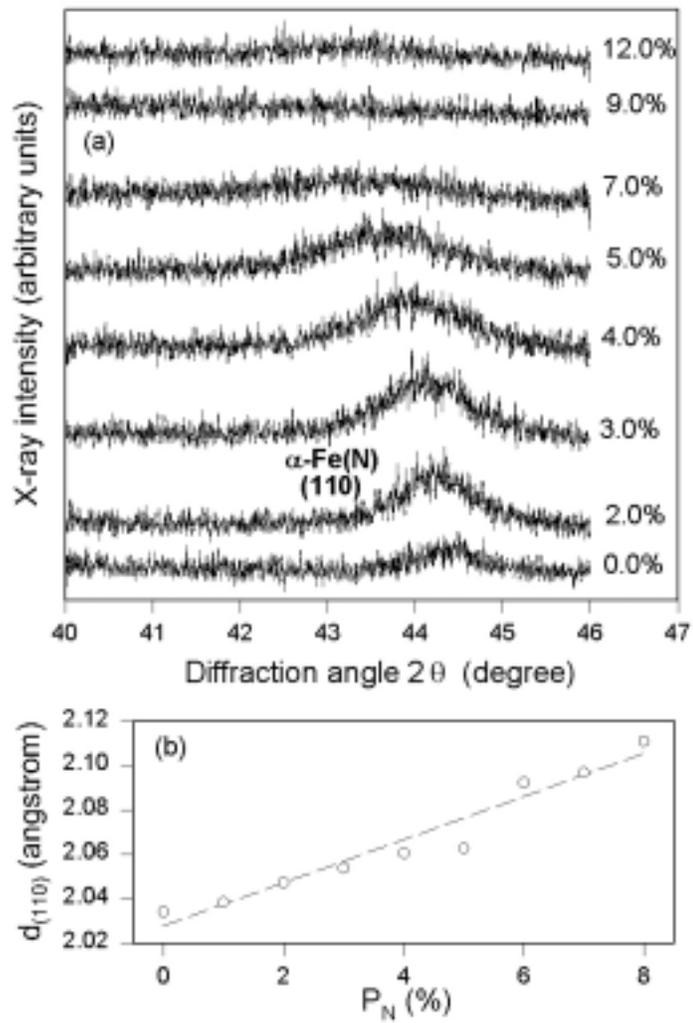

**Figure 1 (a)** XRD step-scan patterns of Fe-N thin films. **(b)** Derived values of the corresponding lattice distant $d_{(110)}$ of the films as a function of $P_N$,

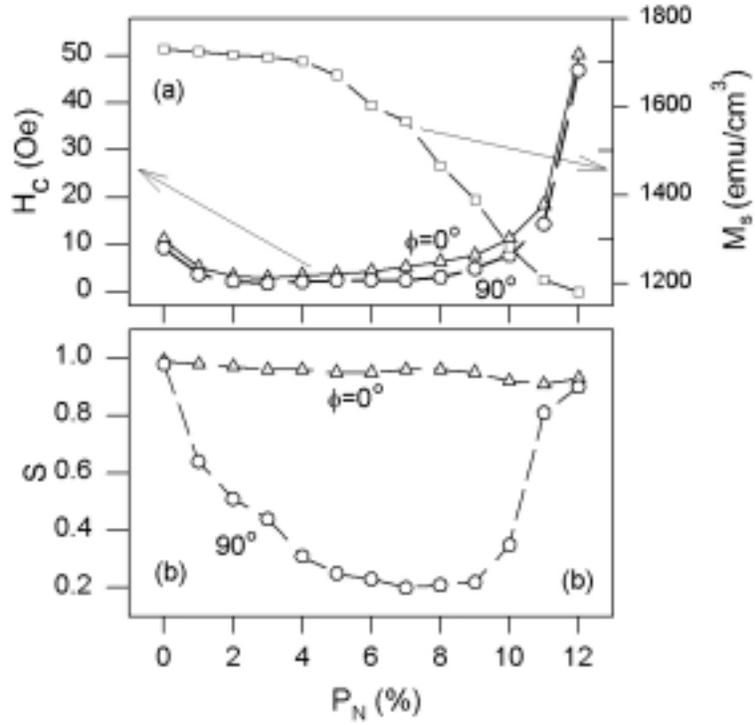

**Figure 2** (**a**) Dependence of magnetic easy ($\phi=0^\circ$) axis coercivity $H_c^e$ and magnetic hard ($\phi=90^\circ$) axis coercivity $H_c^d$, saturation magnetization $M_s$ (**a**), and squareness S (S=$M_r$/$M_s$) (**b**) of as-deposited Fe-N thin films on $P_N$, respectively.

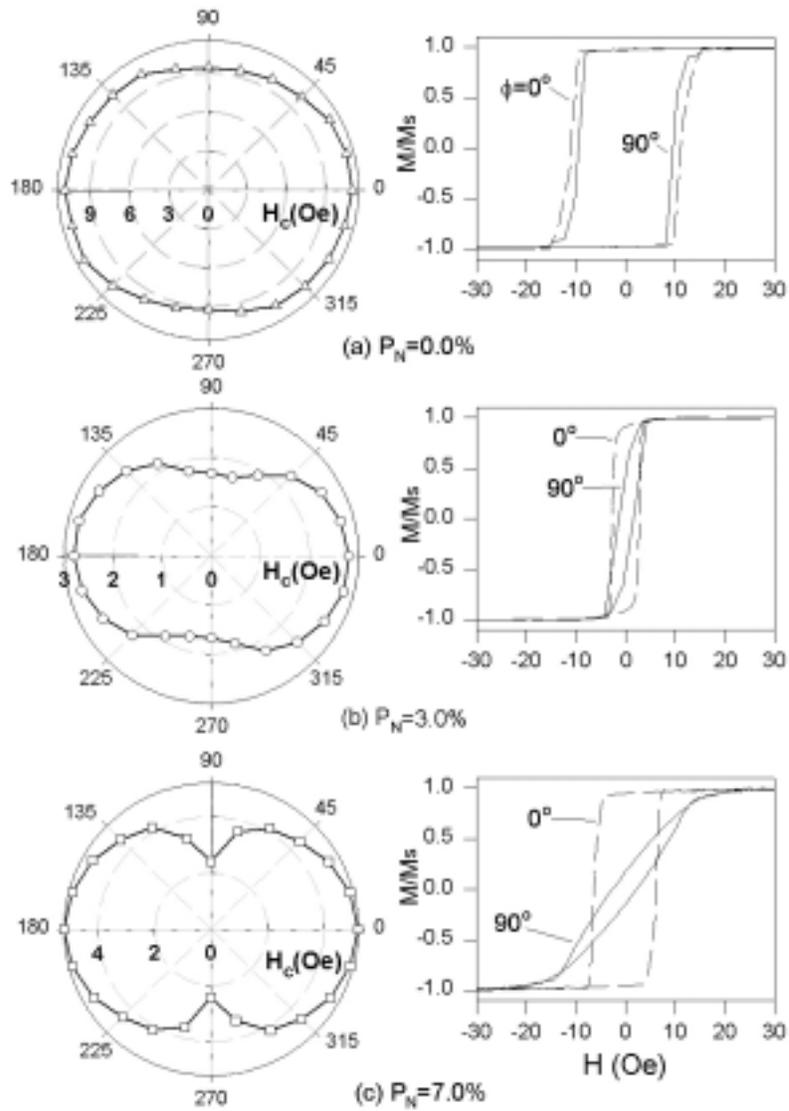

**Figure 3** Polar diagrams of coercivities as a function of angle $\phi$, and Easy ($\phi=0°$) and hard ($\phi=90°$) axis hysteresis loops of Fe-N thin films when $P_N=0.0\%$ (**a**), 3.0% (**b**) and 7.0% (**c**), respectively.

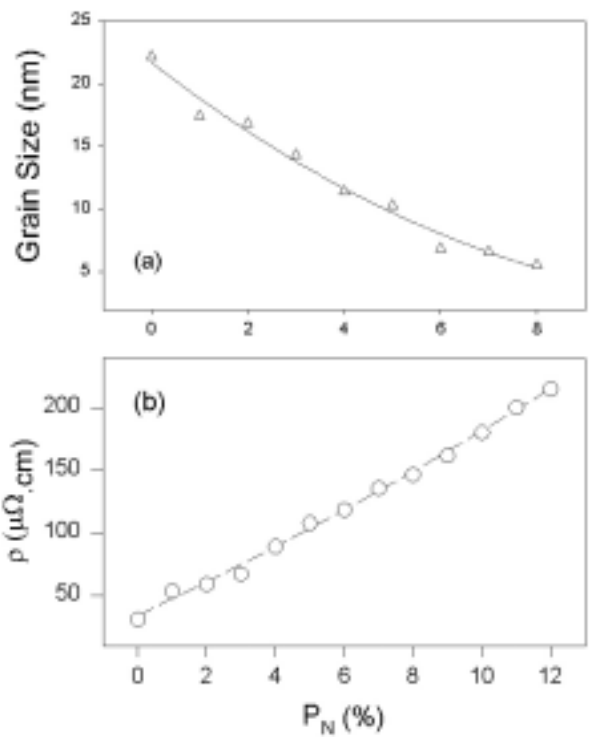

**Figure 4** (a) Average grain sizes calculated from the XRD step-scan patterns according to Scherrer equation as a function of $P_N$. (b) The resistivity $\rho$ of the as-deposited films as a function of $P_N$.

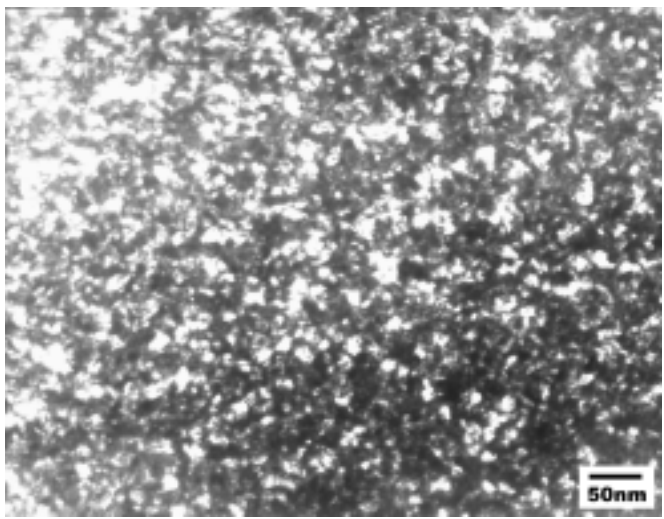

**Figure 5** TEM dark-field plane-view of the film when $P_N=5.0\%$.

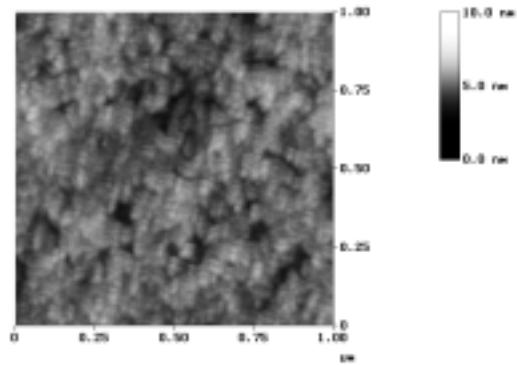

(**a**) P$_N$ = 0.0%

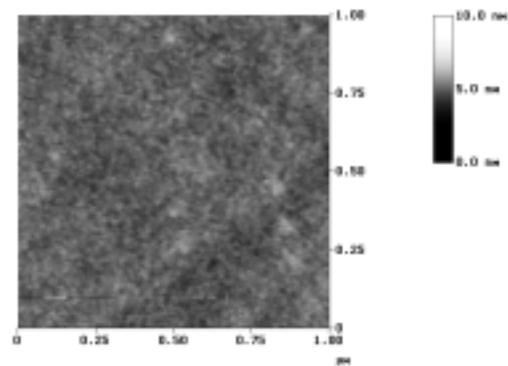

(**b**) P$_N$ = 3.0%

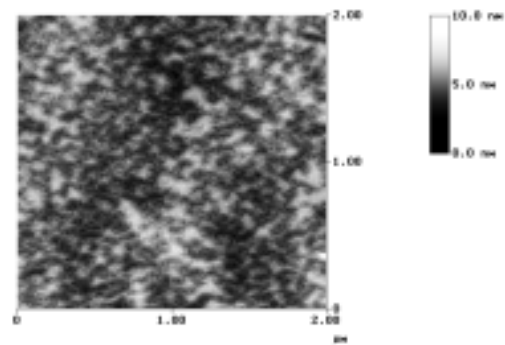

(**c**) P$_N$ = 12.0%

**Figure 6** the AFM tapping mode micrographs of the Fe(N) thin films, corresponding to P$_N$ of 0.0% (**a**), 3.0% (**b**), and 12.0% (**c**), respectively.